\documentclass[12pt,draftcls,onecolumn]{IEEEtran}
\usepackage{graphicx}
\usepackage{psfrag}
\usepackage{amssymb}
\def\be{\begin{equation}}
\def\ee{\end{equation}}
\def\bea{\begin{eqnarray}}
\def\eea{\end{eqnarray}}
\def\beann{\begin{eqnarray*}}
\def\eeann{\end{eqnarray*}}

\def\ns{\hspace{-1mm}}

\def\QED{\mbox{\rule[0pt]{1.5ex}{1.5ex}}}
\def\endproof{\hspace*{\fill}~\QED\par\endtrivlist\unskip}
\newcommand{\real}{{\mathbb{R}}}

\newtheorem{lemma}{Lemma}

\begin{document}
%
%\title{
%A Direct Proof of Theorem 1 in ``Structural Invariant Subspaces of Singular Hamiltonian Systems
%and Nonrecursive Solutions of Finite-Horizon Optimal Control Problems''}
\title{\Large
A Direct Proof of a Theorem Concerning Singular Hamiltonian Systems}
\author{G.~Marro\\[2 mm]
{\small  Dipartimento di Elettronica, Informatica e Sistemistica,
Universit\`a di Bologna}\\
{\small Viale Risorgimento 2, 40136 Bologna - Italy}\\
{\small E-mail: \tt giovanni.marro@unibo.it}\\[2mm]
}
\date{}
\maketitle
\thispagestyle{empty}
%%%%%%%%%%%%%%%%%%%%%%%%%%%%%%%%%%%%%%%%%%%%%%%%%%%%%%%%%%%%%%%%%%%%%%%%%%%%%%%%
%
\begin{abstract}
\noindent
This technical report presents a direct proof of Theorem~1 in \cite{TAC2008-Z-2008} and some consequences that also account for
(20) in \cite{TAC2008-Z-2008}. This direct proof exploits a state space change of basis which replaces the coupled difference equations
(10) in \cite{TAC2008-Z-2008} with two equivalent difference equations which, instead, are decoupled. 
\end{abstract}
%
%%%%%%%%%%%%%%%%%%%%%%%%%%%%%%%%%%%%%%%%%%%%%%%%%%%%%%%%%%%%%%%%%%%%%%%%%%%%%%%%
%
%
%\section{The Direct Proof}
\section{Introduction}
\label{sect_intro}
\vspace{2mm}
Theorem 1 in \cite{TAC2008-Z-2008} provides in (19) the set of the admissible solutions $(x_k,p_k,u_k)$
of the singular  Hamiltonian system (10)
defined on the discrete-time interval $0\le k \le k_f-1$. The proof therein presented is twofold: sufficiency is shown by direct replacement of
(19) in (10); necessity relies on maximality of the involved structural invariant subspaces, as it is deducible from Properties~1 and 2.
In the following, it
will be shown that a direct proof, which does not distinguish between the {\em if} and the {\em only-if} part, but extensively uses relations
pointed out in \cite{TAC2008-Z-2008}, is also feasible. The main point of the direct proof is replacing the coupled diffence equations
(10) in \cite{TAC2008-Z-2008} with two decoupled difference equations by means of a suitable state space basis transformation. The direct proof herein presented can also be used to prove (20) in \cite{TAC2008-Z-2008}, that
expresses the set of the admissible solutions $(x_k.p_k)$ of the same Hamiltonian system in the extended time interval $0\le k \le k_f$.
\section{Direct Proof of Relation (20) and Theorem 1 in~\cite{TAC2008-Z-2008}}
\label{sect_dirproof}
\vspace{2mm}
The direct proof is based on the following lemma.
\vspace{2mm}
\begin{lemma}
\label{lem1}
{\em \ \ The problem of finding the sequences $x_k$, $p_k$, and $u_k$, with $0\,{\le}\,k\,{\le}\,k_f\,{-}\,1$, that solve the 
equations (10) of \cite{TAC2008-Z-2008}, or, equivalently,
\bea
x_{k+1} \ns&\ns = \ns&\ns A\,x_k+B\,u_k,\label{10a} \\
-A^\top p_{k+1} \ns&\ns = \ns&\ns Q\,x_k-p_k+S\,u_k,  \label{10b} \\
-B^\top p_{k+1} \ns&\ns = \ns&\ns S^\top x_k+R\,u_k, \label{10c}
\eea
with $0\,{\le}\,k\,{\le}\,k_f\,{-}\,1$,
can be reduced to that of finding the sequences $v_k$ and $w_k$
that solve the pair of decoupled difference equations
\bea
v_{k+1}\ns&\ns = \ns&\ns A_+\,v_k, \label{20a} \\
A_+^\top\,w_{k+1}\ns&\ns = \ns&\ns w_k, \label{20b} 
\eea
with $0\,{\le}\,k\,{\le}\,k_f\,{-}\,1$,
where $A_+$ is defined by (14) in \cite{TAC2008-Z-2008},
provided that the following correspondences are set up
\bea
x_k \ns&\ns = \ns&\ns v_k+W\,w_k, \label{30a} \\
p_k \ns&\ns = \ns&\ns P_+\,v_k+(-I+P_+W)\,w_k, \label{30b} \\
u_k \ns&\ns = \ns&\ns -K_+v_k + \bar{K}_+w_{k+1} \label{30c},
\eea
where $P_+$ is the positive semidefinite symmetric solution of (11)--(12) in~\cite{TAC2008-Z-2008},
$W$ is the solution of the symmetric discrete Lyapunov equation (15),
$K_+$, and $\bar K_+$ are defined by (13) and (17).} %end em
\end{lemma}
\vspace{2mm}
\IEEEproof
First, the following relation will be shown:
\be
\label{eqW}
-W+B\bar{K}_+ = -A\,W\,A_+^\top.
\ee
\par\noindent
Use of (17) in \cite{TAC2008-Z-2008} yields the identity
\[
  -W+B\bar{K}_+ = -W+ B\,(R+B^\top P_+B)^{-1}(B^\top - B^\top P_+ A\,W A_+^\top - S^\top W A_+^\top)\,=
\]
and, by applying distributivity of the product with respect to the sum,
\[
  =-W + B\,(R+B^\top P_+B)^{-1}B^\top - B\,(R+B^\top P_+B)^{-1} B^\top P_+ A \,W A_+^\top
\]
\[ - B\,(R+B^\top P_+B)^{-1} S^\top W A_+^\top\,=
\] 
and, by collecting $W A_+^\top$ in the last two terms,
\[
  =-W + B\,(R+B^\top P_+B)^{-1} B^\top - B\,(R+B^\top P_+B)^{-1} (B^\top P_+ A + S^\top)\,W A_+^\top\,=
\]
and, by the definition (13) of $K_+$ in \cite{TAC2008-Z-2008}, and summing and subtracting the term $A\,WA_+^\top$
\[
  =-W + B\,(R+B^\top P_+B)^{-1} B^\top - B\,K_+ W A_+^\top + A\,WA_+^\top - A\,WA_+^\top\,=
\]
and, by reordering,
\[
  =(A-BK_+)\,W A_+^\top - W + B\,(R+B^\top P_+B)^{-1} B^\top - A\,WA_+^\top\,=
\]
and, by using (14) in \cite{TAC2008-Z-2008},
\[
  =A_+ W A_+^\top - W + B\,(R+B^\top P_+B)^{-1} B^\top - A\,WA_+^\top\,= 
\]
and, eventually, tacking (15) in \cite{TAC2008-Z-2008} into account,
\[
  =- A\,WA_+^\top\,.
\]
\par
Thus, (\ref{eqW}) is proven. Now we are ready to obtain the difference equation in the unknowns $v_k$ and $w_k$. 
By using (\ref{30a}) and (\ref{30c}) in (\ref{10a}), it follows that:
\[
  v_{k+1}+Ww_{k+1}=Av_k+A\,Ww_k-B\,K_+v_k+B\,\bar{K}_+w_{k+1}\,,
\]
or also
\[
  v_{k+1}=(A-BK_+)\,v_k+(-W+B\bar{K}_+)\,w_{k+1}+A\,Ww_k\,,
\]
or, by the definition (14) in \cite{TAC2008-Z-2008},
\begin{equation}
\label{eqrefx}
  v_{k+1}=A_+v_k+(-W+B\bar{K}_+)\,w_{k+1}+A\,Ww_k\,,
\end{equation}
or, equivalently because of (\ref{eqW}),
\begin{equation}
\label{eqref}
  v_{k+1}=A_+v_k-A\,WA_+^\top w_{k+1}+A\,Ww_k\,.
\end{equation}
\par 
Similarly, by using (\ref{30a})--(\ref{30c}) in (\ref{10b}), the following is obtained:
\[
  -A^\top\bigl(P_+ v_{k+1} + (P_+ W - I)\,w_{k+1}\bigr) = 
\]
\[
 =Q (v_k + W w_k) - \bigl(P_+v_k - (P_+ W - I) w_k\bigr) 
  + S (-K_+ v_k + \bar{K}_+ w_{k+1})\,,
\]
or
\[
  -A^\top P_+ v_{k+1} - A^\top (P_+ W - I)\,w_{k-1} =
\]
\[= Q v_k +QWw_k - P_+ v_k - (P_+ W - I)w_k -S K_+ v_k + S \bar{K}_+ w_{k+1}\,.
\]
\par
By the identity $- A^\top (P_+ W - I) = Q W A_+^\top - (P_+ W - I) A_+^\top + S \bar{K}_+$ (see the proof of Property 2 in \cite{TAC2008-Z-2008} -- second row block), the following holds:
\[
  -A^\top P_+ v_{k+1} + \bigl(QWA_+^\top - (P_+ W - I) A_+^\top + S \bar{K}_+\bigr)w_{k+1}= 
\]
\[
 = Qv_k + QWw_k -P_+v_k - (P_+ W - I) w_k - SK_+ v_k + S \bar{K}_+ w_{k+1}\,,
\]
and, by doing away with the terms $S \bar{K}_+ w_{k+1}$ at the right of both members,
\[
  -A^\top P_+ v_{k+1} + \bigl( QW - (P_+ W - I) \bigr) A_+^\top w_{k+1}=
\] 
\[ =(Q-P_+-SK_+)v_k +\bigl(QW - (P_+ W - I)\bigr) w_k\,. 
\]
\par
Recall the identity $Q-P_+-SK_+ = -A^\top P_+ A_+$ (see the proof of Property 1 in \cite{TAC2008-Z-2008} -- second row block), the following is obtained:
\be
\label{eqdot}
   -A^\top P_+ v_{k+1} + \bigl(QW - (P_+W-I) \bigr) A_+^\top w_{k+1}=
   -A^\top P_+A_+ v_k + \bigl( QW - (P_+ W - I) \bigr) w_k\,.
\ee
Let us multiply both members of (\ref{eqref}) by $A^\top P_+$, thus obtaining
\be
\label{eqtridot}
  A^\top P_+ v_{k+1}=A^\top P_+ A_+ v_k-A^\top P_+ A\,WA_+^\top w_{k+1}+A^\top P_+A\,Ww_k\,,
\ee
and, by summing both members of (\ref{eqdot}) and (\ref{eqtridot}), it follows that
\[
  \bigl(QW - (P_+ W - I) \bigr) A_+^\top w_{k+1} =
\]
\[ = \bigl(QW - (P_+ W - I) \bigr) w_k - A^\top P_+ A W A_+^\top w_{k+1} + A^\top P_+ A W  w_k\,.
\]
By collecting $w_{k+1}$ on the left and $w_k$ on the right, it follows that
\[
  \bigl(QW - (P_+ W - I) + A^\top P_+ AW \bigr) A_+^\top w_{k+1} =  \bigl(QW - (P_+ W - I) + A^\top P_+ AW \bigr) w_k\,,
\]
or $A_+^\top w_{k+1} = w_k$, that is (\ref{20b}).
Taking into account this latter equation in (\ref{eqref}) one gets
\[
  v_{k+1} = A_+ v_k - AWw_k + AWw_k\,,
\]
or $v_{k+1}=A_+v_k$, that is  (\ref{20a}). \endproof
\vspace{4mm}
Now we are ready to conclude the direct proof of both (20) and (19) in \cite{TAC2008-Z-2008}.
Refer to the pair of decoupled difference equations (\ref{20a}), (\ref{20b}), defined in the
time interval $0\le k \le k_f-1$. Their solutions can be expressed as
\be
\label{eqf1}
\begin{array}{rcl}
v_k \ns&\ns = \ns&\ns A_+^k \alpha\,, \\ [2mm]
w_k \ns&\ns = \ns&\ns (A_+^\top)^{k_f-k} \beta\,,
\end{array}
\quad 0 \le k \le k_f\,,
\ee
where $\alpha,\beta\in\real^n$ are parameters. Substitution of (\ref{eqf1}) in (\ref{30a}),
(\ref{30b}) yields
\[
\begin{array}{rcl}
x_k \ns&\ns = \ns&\ns A_+^k \alpha+W\,(A_+^\top)^{k_f-k} \beta\,, \\ [2mm]
p_k \ns&\ns = \ns&\ns P_+ A_+^k \alpha +(P_+W-I)(A_+^\top)^{k_f-k} \beta\,,
\end{array}
\quad 0 \le k \le k_f\,,
\]
that, re-written in compact notation as
\[
\left[\begin{array}{c} x_k \\ p_k \end{array}\right]=\left[\begin{array}{c}
I \\ P_+ \end{array} \right]\, A_+^k \alpha + \left[\begin{array}{c} W \\
P_+W-I \end{array}\right] (A_+^\top)^{k_f-k}\beta\,, \quad 0\le k \le k_f\,,
\]
coincides with equation (20) in \cite{TAC2008-Z-2008}. \\ 
\par
To prove equation (19) in \cite{TAC2008-Z-2008}, let us substitute (\ref{20b}), i.e.,
\[
  w_k=A_+^\top\,w_{k+1}\,, \quad 0\le k \le k_f-1\,,
\]
in (\ref{30a}), (\ref{30b}), thus obtaining
\be
\label{eqxx}
\begin{array}{rcl}
x_k \ns&\ns = \ns&\ns v_k + W\,A_+^\top w_{k+1}\,, \\ [2mm]
p_k \ns&\ns = \ns&\ns (P_+W-I)\,A_+^\top w_{k+1}\,,
\end{array}
\quad 0\le k \le k_f-1\,.
\ee
Using (\ref{eqf1}) in (\ref{eqxx}) yields
\be
\label{eqn45}
\begin{array}{rcl}
  x_k \ns&\ns = \ns&\ns A_+^k\,\alpha + W A_+^\top (A_+^\top)^{k_f-k-1}\beta\,, \\ [2mm]
  p_k \ns&\ns = \ns&\ns P_+\,A_+^k\,\alpha + (P_+W-I) A_+^\top (A_+^\top)^{k_f-k-1}\beta\,,
\end{array}
\quad 0 \le k \le k_f-1\,,
\ee
while using (\ref{eqf1}) in (\ref{30c}) provides
\be
\label{eqn6}
  u_k=-K_+ A_+^k\,\alpha + \bar{K}_+(A_+^\top)^{k_f-k-1}\beta\,,
  \quad 0 \le k \le k_f-1\,.
\ee
Equations (\ref{eqn45}), (\ref{eqn6}) can be re-written in compact form as
\[
\left[\begin{array}{c} x_k \\ p_k \\ u_k \end{array}\right] =
\left[\begin{array}{c} I \\ P_+ \\ -K_+ \end{array}\right] A_+^k\, \alpha +
\left[\begin{array}{c} W\,A_+^\top \\ (P_+W-I)A_+^\top \\ \bar{K}_+ \end{array}\right]
(A_+^\top)^{k_f-k-1}\beta\,,
\quad 0 \le k \le k_f-1\,,
\]
that coincides with (19) in \cite{TAC2008-Z-2008}. Thus, Theorem 1 in \cite{TAC2008-Z-2008} has
been directly proven by using the correspondences stated in Lemma~\ref{lem1}.
\vspace{2mm}
\bibliographystyle{IEEEtran}
\bibliography{remarksbib}
\end{document}